\documentclass[conference,9pt]{IEEEtran}

\usepackage{cite}
\usepackage{amsmath,amssymb,amsfonts}
\usepackage{algorithmic}
\usepackage{graphicx}
\usepackage{textcomp}
\usepackage{xcolor}
\def\BibTeX{{\rm B\kern-.05em{\sc i\kern-.025em b}\kern-.08em
    T\kern-.1667em\lower.7ex\hbox{E}\kern-.125emX}}

\usepackage{booktabs} 
\usepackage{soul}
\graphicspath{{figs/}}
\usepackage{braket}

\usepackage{etoolbox}
\makeatletter
\patchcmd{\@maketitle}
  {\addvspace{0.5\baselineskip}\egroup}
  {\addvspace{-0.5\baselineskip}\egroup}
  {}
  {}

\begin{document}

\title{\huge{DeepQMLP: A Scalable Quantum-Classical Hybrid Deep \\Neural Network Architecture for Classification}}

\author{\IEEEauthorblockN{Mahabubul Alam}
\IEEEauthorblockA{\textit{Department of Electrical Engineering} \\
\textit{Penn State University}\\
mxa890@psu.edu}
\and
\IEEEauthorblockN{Swaroop Ghosh}
\IEEEauthorblockA{\textit{Department of Electrical Engineering} \\
\textit{Penn State University}\\
szg212@psu.edu}
}

\maketitle

\begin{abstract}

Quantum machine learning (QML) is promising for potential speedups and improvements in conventional machine learning (ML) tasks (e.g., classification/regression). The search for ideal QML models is an active research field. This includes identification of efficient classical-to-quantum data encoding scheme, construction of parametric quantum circuits (PQC) with optimal expressivity and entanglement capability, and efficient output decoding scheme to minimize the required number of measurements, to name a few. However, most of the empirical/numerical studies lack a clear path towards scalability. Any potential benefit observed in a simulated environment may diminish in practical applications due to the limitations of noisy quantum hardware (e.g., under decoherence, gate-errors, and crosstalk). We present a scalable quantum-classical hybrid deep neural network (DeepQMLP) architecture inspired by classical deep neural network architectures. In DeepQMLP, stacked shallow Quantum Neural Network (QNN) models mimic the hidden layers of a classical feed-forward multi-layer perceptron network. Each QNN layer produces a new and potentially rich representation of the input data for the next layer. This new representation can be tuned by the parameters of the circuit. Shallow QNN models experience less decoherence, gate errors, etc. which make them (and the network) more resilient to quantum noise. We present numerical studies on a variety of classification problems to show the trainability of DeepQMLP. We also show that DeepQMLP performs reasonably well on unseen data and exhibits greater resilience to noise over QNN models that use a deep quantum circuit. DeepQMLP provided up to 25.3\% lower loss and 7.92\% higher accuracy during inference under noise than QMLP.

\end{abstract}

\section{Introduction}
Quantum computing is one of the major transformative technologies. Although quantum computing is still in a nascent stage, the community is seeking computational advantages with quantum computers (i.e., quantum supremacy) for practical applications. Recently, Google claimed quantum supremacy with a 53-qubit quantum processor to complete a computational task in 200 seconds that might take 10K years \cite{arute2019quantum} (later rectified to 2.5 days \cite{pednault2019quantum}) on the state-of-the-art supercomputers. This experiment was a significant milestone for quantum computing even though the computational task used for this experiment had no practical value.

\begin{figure*}[t]
 \begin{center}
    \includegraphics[width=0.95\textwidth]{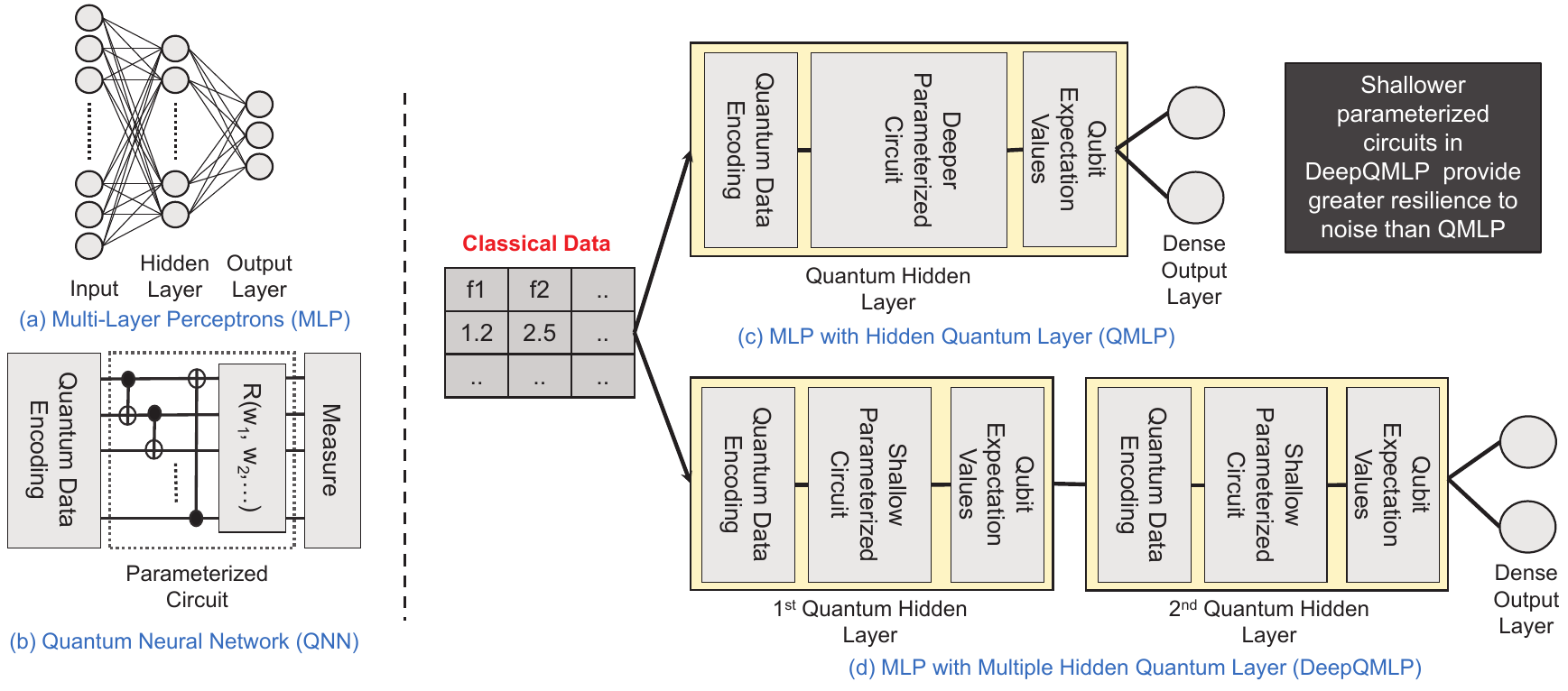}
 \end{center}
 \vspace{-4mm}
\caption{Conventional Multi-Layer Perceptron (MLP) and Quantum Neural Network (QNN) architectures (a)-(b), alongside the proposed Quantum Multi-Layer Perceptrons (QMLP) and Deep Quantum Multi-Layer Perceptrons (DeepQMLP) architectures (c)-(d). While QMLP uses a deeper parameterized circuit to accommodate larger search space, DeepQMLP uses multiple shallow-depth circuits. Shallower circuits provide greater robustness against quantum noises (less accumulation of gate errors and decoherence over each circuits) to DeepQMLP over QMLP. 
} 
\vspace{-4mm}
\label{fig:idea}
\end{figure*}

Quantum machine learning (QML) is a promising application domain to archive quantum advantage with noisy quantum computers in the near term. Numerous QML models built upon parametric quantum circuits (PQC), also referred to as quantum neural networks (QNN), are already available in the literature \cite{farhi2018classification, killoran2019continuous, cong2019quantum}. A PQC is a quantum circuit with tunable parameterized gates as shown in Fig. \ref{fig:idea}(b) (w1,w2,... are the tunable parameters). A PQC may generate various output states based on the values of these parameters. QNN models are claimed to be more expressive compared to the classical neural networks \cite{du2020expressive, wright2020capacity, abbas2020power}. In other words, QNN models have a higher capability to approximate the desired functionality (e.g., classifying data samples) compared to the classical models of a similar scale (e.g., with the same number of tunable parameters/weights). Deep Neural Networks (DNN) have experienced huge success in machine learning (ML) in the past decade (essentially superseding most other models) because they are powerful function approximators. With an even higher ability to approximate functions, QNN holds great potential for the future. 


A conventional QNN architecture is shown in Fig. \ref{fig:idea}(b). In a typical QNN model, the input data is encoded into a quantum state using a suitable encoding scheme (e.g., angle encoding, amplitude encoding, etc.) \cite{schuld2020circuit}. The encoding is followed by layers of PQC with tunable parameters (w1,w2,... in Fig. \ref{fig:idea}(b)). These parameters are analogous to weights in a classical neural network. In the end, the output quantum state of the PQC is measured (/sampled) on the appropriate basis (e.g., the default Pauli-Z measurement basis in IBM quantum computers \cite{cross2018ibm}). The sampling process is repeated many times with the same parameters. A cost function is derived from the measurements. A classical optimizer (e.g., gradient-descent) updates the parameter values to minimize the cost. 

The choice of the PQC can have a significant impact on the performance (e.g., trainability) of a QNN model. For instance, deep PQC with lots of parameters may be desirable for learning but may experience vanishing gradient problems (also referred to as barren plateaus) making it harder for the gradient-based optimizers to navigate through the solution space \cite{mcclean2018barren}. 
Moreover, quantum computers are plagued with various noise sources such as gate error, readout error, decoherence, and crosstalk \cite{alam2020design}. The output quantum state can be random (i.e., meaningless) if the noise accumulation is high. A large amount of noise can also induce a barren plateau in the QNN solution space \cite{wang2021noise}.

Shallow-depth circuits are preferred for QNN to avoid the aforementioned issues \cite{mcclean2018barren, cerezo2021cost}. However, shallow-depth circuits may often be unable to approximate complex functionality (similar to shallow classical neural networks with small number of parameters). 

In this article, we propose two new quantum-classical hybrid deep neural network architectures: Quantum Multi-Layer Perceptrons (QMLP) and DeepQMLP to partially address the aforementioned issues. 
Both architectures are inspired by conventional Multi-Layer Perceptron (MLP) networks used in deep learning. In MLP, multiple layers of neurons are used to define and search through a solution space for a given ML task (Fig. \ref{fig:idea}(a)). Neurons of successive layers are connected through trainable weights. The first and the last layers of MLP are commonly referred to as input and output layers. The internal layers are referred to as hidden layers. Typically, MLP models contain multiple hidden layers. In QMLP, the hidden layer of an MLP is mimicked by a QNN layer as shown in Fig. \ref{fig:idea}(c). The QNN takes a quantum encoded representation of the classical data and produces an output representation (e.g., Pauli-Z expectation values of the qubits) which is fed to the classical output layer. The network can be trained with any conventional loss function. However, in this work, we only use cross-entropy loss. In DeepQMLP, multiple shallow-depth QNN models (two used in this work) are used as hidden layers of an MLP (Fig. \ref{fig:idea}(d)). Each layer produces a new representation for the next layer. For example, the qubit expectation values of the first hidden layer in Fig. \ref{fig:idea}(d) are used as the inputs to the second hidden layer. 

QMLP uses a deep QNN alongside a classical dense output layer to exploit the higher expressive power of QNN. To accommodate a larger search space, the quantum hidden layer in QMLP needs a deep parameterized circuit. However, deep circuits are more error-prone. DeepQMLP addresses the issue by using a series of shallow-depth quantum circuits stacked one after another. The shallower circuits require fewer number of gates and execution time which reduces the accumulation of gate errors and decoherence. Thus, the architecture shows more robustness against noise.

\textbf{Contributions:} We, (a) present two new quantum-classical hybrid neural network architectures (QMLP and DeepQMLP) for classification, (b) show the trainability of the proposed models through numerical studies with 4 synthetic datasets and the `iris' dataset across 78 training runs with varying depth of the parametric quantum circuits, and (c) present an empirical proof-of-concept study to exhibit greater noise resilience of the DeepQMLP architecture.    



\section{Preliminaries} \label{basics}

\noindent {\bf{Qubits and Quantum Gates:}} Qubit is analogous to classical bits however, a qubit can be in a superposition state i.e., a combination of $\ket{0}$ and $\ket{1}$ at the same time. Quantum gates such as single qubit (e.g., Pauli-X ($\sigma_x$) gate) or multiple qubit (e.g., 2-qubit CNOT gate) gates modulate the state of qubits. These gates can perform a fixed computation (e.g. an X gate flips a qubit state) or a computation based on a supplied parameter (e.g. the RY($\theta$) gate rotates the qubit along the Y-axis by $\theta$). A two-qubit gate changes the state of one qubit (commonly referred to as the target qubit) based on the current state of the other qubit (commonly referred to as the control qubit). For example, The CNOT gate flips the target qubit if the control qubit is in $\ket{1}$ state. Similarly, the CRZ($\theta$) gate rotates the target qubit along Z-axis by $\theta$ if the control qubit is in $\ket{1}$ state. 

\noindent {\bf{Noise in Quantum Computation:}} Quantum gates are error-prone. Besides, qubits suffer from decoherence i.e., they spontaneously interact with the environment and lose states. Therefore, the output of a quantum circuit is erroneous. The deeper quantum circuit needs more time for execution and gets affected by decoherence. More gates in the circuit also increase the accumulation of gate error. Thus, lower depth and the number of gates in the circuit improve noise resiliency. Parallel gate operations on different qubits can affect each other's performance which is known as crosstalk. 


\noindent {\bf{Expectation Value of an Operator:}} Expectation value is the average of the eigenvalues, weighted by the probabilities that the state is measured to be in the corresponding eigenstate. Mathematically, expectation value of an operator ($\sigma$) is defined as $\langle\psi|\sigma|\psi\rangle$ where $|\psi\rangle$ is the qubit state vector. For example, the expectation value of the Pauli-Z operator ($\sigma_z$) is $\langle\psi|\sigma_z|\psi\rangle$. If a qubit yields more $\ket{0}$ $(\ket{1})$ than $\ket{1}$ $(\ket{0})$, its Pauli-Z expectation value will be positive (negative). This value will vary in the range [-1, 1].

\begin{figure} [t]
 \begin{center}
    \includegraphics[width=0.48\textwidth]{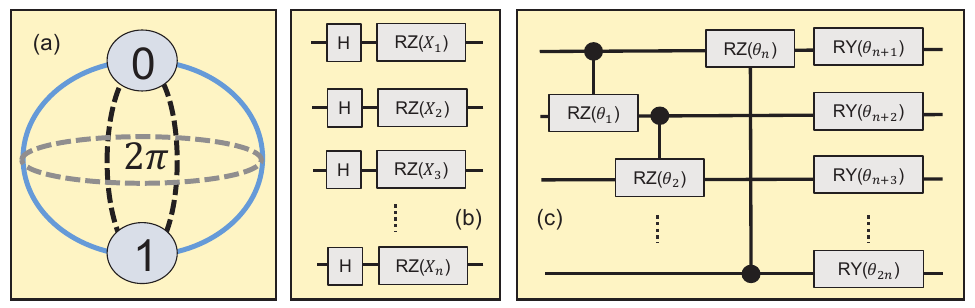}
 \end{center}
 \vspace{-4mm}
\caption{(a) The Bloch sphere representation of a qubit. From any given point, a qubit can be rotated along the X, Y, or Z-axis. In 2$\pi$ intervals, the states will repeat, (b) angle encoding used in this work, and (c) parametric layer structure used in this work.} 
\vspace{-4mm}
\label{fig:elements}
\end{figure}

\noindent {\bf{Quantum Neural Network:}} QNN involves parameter optimization of PQC to obtain a desired input-output relationship. The PQC generally consists of three segments: (i) a classical to quantum data encoding (also referred to as embedding in the literature) circuit, (ii) parameterized circuit, and (iii) measurement operations. A variety of encoding methods are available in the literature \cite{schuld2021effect}. For continuous variables, the most widely used encoding scheme is angle encoding \cite{abbas2020power, schuld2020circuit, schuld2021effect, lloyd2020quantum} where a continuous variable input classical feature is encoded as a rotation of a qubit along the desired axis (X/Y/Z). For `n' classical features, we require `n' qubits. In this work, we use RZ(X1) on a qubit in superposition to encode any classical feature `X1' as shown in Fig. \ref{fig:elements}(b). The H (Hadamard) gate is used to put a qubit in superposition. As the states produced by a qubit rotation along any axis will repeat in 2$\pi$ intervals (Fig. \ref{fig:elements}(a)), features are generally scaled within 0 to 2$\pi$ in a data pre-processing step. One can restrict the values between -$\pi$ to $\pi$ to accommodate features with both negative and positive values. 

The parametric circuit has two components: entangling operations and parameterized single-qubit rotations. The entanglement operations are a set of two-qubit (or more) operations between pairs of qubits to generate correlated states \cite{lloyd2020quantum}. The following parametric single-qubit operations are used to define and search through the solution space. Note that, one can also use parametric two-qubit operations in entanglement to expand the search space. This combination of entangling and single-qubit rotation operations is referred to as a parametric layer in QNN. We use the parametric layer shown in Fig. \ref{fig:elements}(c) throughout this work. Here, we use qubit pair-wise CRZ($\theta$) gates to create the entanglement followed by rotations along Y-axis using RY($\theta$) operations. Normally, these layers are repeated multiple times to extend the search space of QNN \cite{schuld2020circuit, abbas2020power}. The measurement operations are chosen to accommodate the desired cost function.

\begin{figure}
 \begin{center}
    \includegraphics[width=0.4\textwidth]{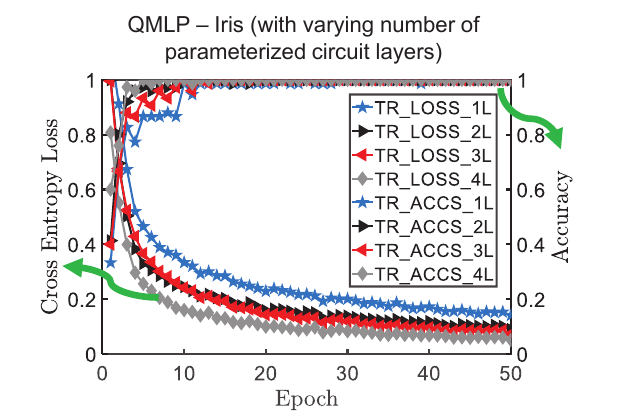}
 \end{center}
 \vspace{-4mm}
\caption{Performance (loss and accuracy) of QMLP on the 'Iris' dataset over 50 epochs with 1, 2, 3, and 4 parametric layers. Performance improves with added layers (lower loss/higher accuracy after the same number of epochs).} 
\vspace{-4mm}
\label{fig:added_layer}
\end{figure}

\begin{figure*} [t]
 \begin{center}
    \includegraphics[width=0.95\textwidth]{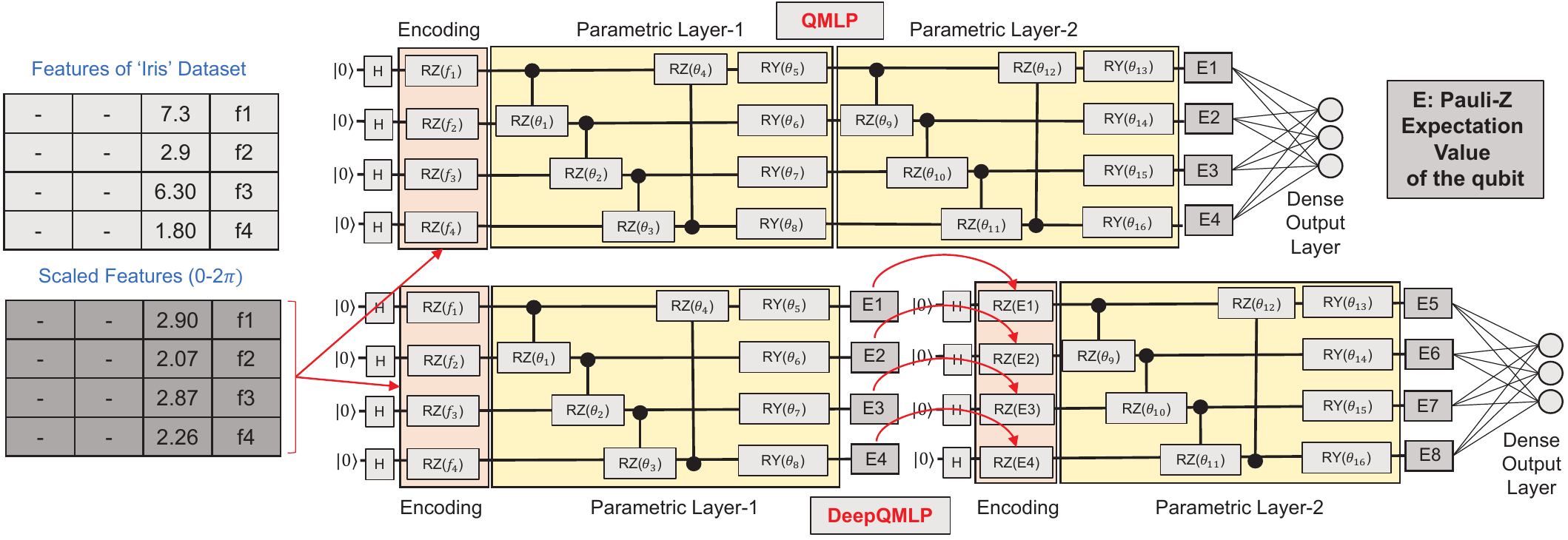}
 \end{center}
 \vspace{-4mm}
\caption{The QMLP and DeepQMLP network architectures used in this work (with 2 parametric layers) for 'Iris' dataset classification. (f1, f2, f3, f4) are the 4 features of the dataset. Both these networks require 4 qubits. The number of trainable parameters i.e., $\theta_1$, ..., $\theta_{16}$ in these two networks are identical. The final layer is a classical dense layer (3 neurons) with SoftMax activation.}
\vspace{-4mm}
\label{fig:iris_ckt}
\end{figure*}

\section{Proposed Architectures} \label{proposals}



\begin{figure}
 \begin{center}
    \includegraphics[width=0.48\textwidth]{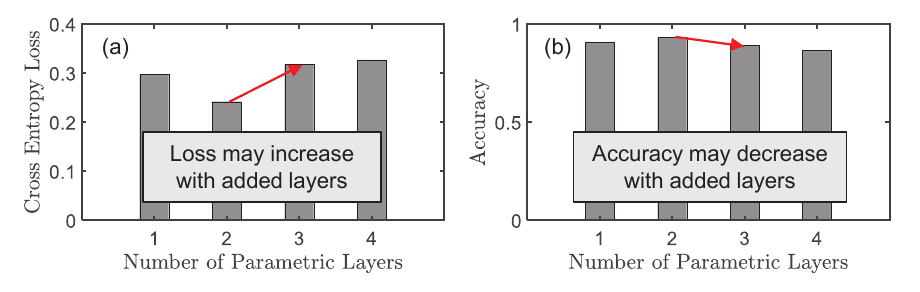}
 \end{center}
 \vspace{-4mm}
\caption{(a) Loss and (b) accuracy of the trained QMLP models ('Iris' dataset with 1, 2, 3, and 4 parametric layers) on the training data during inference under noise (using noise parameters of IBM Melbourne device). The performance may decrease (higher loss/lower accuracy) with added layers under noise. A larger accumulation of noise in deeper circuits corrupts the output quantum state significantly which causes performance degradation.} 
\label{fig:iris_inference}
\vspace{-4mm}
\end{figure}

\begin{figure*} 
 \begin{center}
    \includegraphics[width=0.95\textwidth]{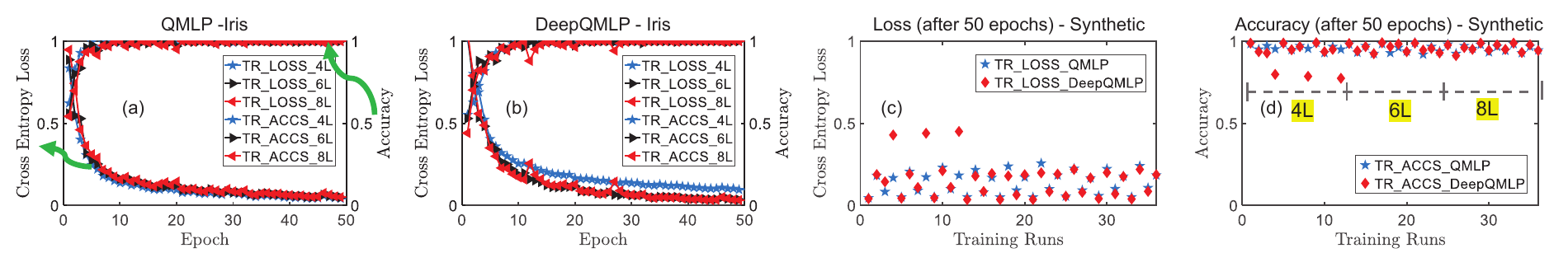}
 \end{center}
 \vspace{-4mm}
\caption{Loss and accuracy over 50 epochs of training ('Iris' dataset) with 4, 6, and 8 parametric layers with the (a) QMLP, and (b) DeepQMLP architectures; (c) loss and (d) accuracy - at the end of the 36 training runs on the synthetic datasets with the QMLP and DeepQMLP architectures with 4, 6, and 8 parametric layers. In all the training runs, both QMLP and DeepQMLP networks achieved a small cross-entropy loss and high accuracy over the training dataset which indicates that these networks are trainable. DeepQMLP performance is mostly at the same level with QMLP (training under zero noise).} 
\vspace{-4mm}
\label{fig:training}
\end{figure*}

\subsection{QMLP}

QMLP has two major components: (i) Hidden Quantum Layer and (ii) Classical Dense Output Layer. Structurally, the hidden quantum layer is similar to the conventional QNN. However, unlike conventional QNN, which uses the measurement operations as a direct component of the cost function, this hidden quantum layer produces a representation of the data which is used as an input to the classical dense layer. The hidden layer outputs and the classical layers are connected by trainable weights that further expand the search space. QMLP utilizes the higher expressive power of the hidden quantum layer to build powerful hybrid classification models. 

\noindent {\bf{Hidden Quantum Layer:}} The hidden quantum layer starts with angle encoding of the classical feature variables. Each feature is scaled within 0 to 2$\pi$ (/-$\pi$ to $\pi$). We use rotation along the Z-axis on a qubit which is in a superposition state (created using Hadamard gate prior to the rotation) as our preferred angle encoding method \cite{abbas2020power}. Note that one can use other encoding methods in QMLP as well. The encoding is followed by a multi-layer parametric circuit. We use pairwise CRZ($\theta$) operations to create the entanglement followed by parameterized rotations of the qubits along the Y-axis using RY($\theta$) operations. Again, any suitable parametric circuit structure may be used. The output from the hidden quantum layer is taken as the Pauli-Z expectation values of all the qubits. This hidden layer requires `n' qubits for a classification problem with `n' continuous feature variables. It also produces `n' output features which are fed to the densely connected output layer. The number of trainable parameters in our chosen hidden layer is `2*n*L' where L is the number of parametric circuit layers. 

\noindent {\bf{Classical Dense Output Layer:}} The output layer is a classical dense layer. The number of neurons is equal to the number of classes in the given problem. 
These neurons are densely connected with the hidden quantum layer outputs. For a classification problem with `n' features and `m' classes, there are `n*m' trainable weights between the hidden layer and the output layer. The output neurons use SoftMax activation to produce the class probabilities. 

\noindent {\bf{Training QMLP:}} The network can be trained using a conventional mini-batch gradient descent approach with any suitable optimizer (e.g., Adam, Adagrad, etc.). The output of the dense layer is used to calculate the loss of the network for a batch of samples using any suitable loss function (e.g., MSE, MAE, Cross-entropy Loss, etc.). The loss can be backpropagated using the backpropagation algorithm to compute gradients and update the trainable weights/parameters in the network using gradient-based optimization algorithms. Multiple methods exist to calculate the gradient of a quantum circuit output with respect to the parameters such as parameter-shift rule, finite difference, adjoint, backprop, etc. \cite{bergholm2018pennylane}. Parameter-shift rule or finite difference methods rely upon repeated evaluation of the outputs with slightly altered parameters. Although adjoint, backprop, etc. methods are much faster than finite difference/parameter-shift rule, they are not suitable for hardware as they require the knowledge of intermediate states of the circuit which is only accessible during simulation. In this work, we use the adjoint method \cite{bergholm2018pennylane}.

{\bf{Example 1:}} An example of the QMLP architecture (with two parametric layers) is shown in Fig. \ref{fig:iris_ckt} to classify the `Iris' dataset which contains 150 samples that belong to 3 different classes. The dataset has 4 features. These features are continuous variables that are scaled within 0 to 2$\pi$. The hidden quantum layer uses 4 qubits. The output dense layer has 3 neurons. Overall, the network has (2x4x2 + 4x3) or 28 trainable parameters/weights (16 circuit parameters and 12 classical weights). In Fig. \ref{fig:added_layer}, we show the training loss and accuracy of the dataset over 50 epochs of training using a different number of circuit layers (referred to as 1L, 2L, 3L, and 4L; 1L corresponds to 1 parametric circuit layer). Cross-entropy loss is used for the loss function, Adagrad with a learning rate of 0.5 is used as an optimizer. Quantum noise is not considered. Note that the learning improved with an increasing number of layers as evident from Fig. \ref{fig:added_layer} (achieving lower loss/higher accuracy at the same number of epochs). For example, the loss was 56.3\% lower and the accuracy was 1.35\% higher with 4L compared to 1L after 20 epochs. However, increasing the number of layers beyond 4 yielded diminishing returns as evident from Fig. \ref{fig:training}(a). 

\subsection{DeepQMLP}

Note that the performance improvement with an increasing number of layers in QMLP may not hold true under noise. In reality, deeper circuits are more susceptible to quantum noise due to higher gate error accumulation and decoherence. A circuit with 4 layers has twice as many gates and requires twice as much execution time as a 2L circuit. Therefore, the output state will be considerably more erroneous in a 4L circuit compared to a 2L circuit.

For further illustration, we took the trained QMLP models for `Iris' classification with 1, 2, 3, and 4 layers (Fig. \ref{fig:added_layer}) and measured the loss and accuracy over the entire training dataset using the noise parameters of IBM Melbourne device (simulated using Qiskit \cite{cross2018ibm}). The results are shown in Fig. \ref{fig:iris_inference}(a)-(b). Note that, even though, the training produced a lower loss and higher accuracy with an increasing number of layers (from 1 to 4 in Fig. \ref{fig:added_layer}), during inference under noise (which is a reasonable use case since the models are expected to be used in actual noisy quantum hardware) QMLP with a higher number of layers performed poorly. For example, QMLP with 4 layers showed 8.99\% higher loss and 4.62\% lower accuracy than the 1 layer model. As a potential solution to address this limitation of QMLP, we propose DeepQMLP where we use multiple shallow-depth quantum circuits stacked one after another instead of a deeper circuit. This architecture is analogous to classical MLP with multiple hidden layers. Here, the classical hidden layers are mimicked by shallow-depth PQC.

\noindent {\bf{Hidden Layers in DeepQMLP:}} Each hidden layer of DeepQMLP produces a representation of the data that is fed to the next layer in the network. The first hidden layer in DeepQMLP is identical to the hidden layer in QMLP. However, instead of feeding the output expectation values to a densely connected classical layer, these outputs are again encoded as a quantum state in a subsequent PQC using angle encoding. This PQC produces another set of outputs that are encoded as the quantum state of the next PQC (if there are more than two hidden quantum layers). The last hidden quantum layer is connected to a densely connected classical layer similar to QMLP. The circuit structure of these hidden layers can be different from each other. However, in this work, we use identical circuit structures for all the hidden layers. The training procedure is similar to QMLP. 

{\bf{Example 2:}} An example of the DeepQMLP architecture 
is shown in Fig. \ref{fig:iris_ckt} to classify the `Iris' dataset. The 4 input classical features are encoded into the qubits of the first hidden layer using angle encoding. The four expectation values of the qubits from the output of the first hidden layer (E1, E2, E3, and E4) are encoded as a quantum state in the next hidden layer using angle encoding. The output of the second hidden layer feeds the classical dense layer with 3 neurons. Overall, the network has (2*4*1*2 + 4*3) or 28 parameters (16 circuit parameters and 12 classical weights). In Fig. \ref{fig:training}(b), we show the training loss and accuracy of DeepQMLP over 50 epochs of training (with 4, 6, and 8 cumulative circuit layers). We show the performance of QMLP with a similar number of circuit parameters in Fig. \ref{fig:training}(a). Note that, the performance (noiseless) of the DeepQMLP models stayed close to the QMLP models with the same number of parameters (except the 4 layer one). For example, the difference in training loss and accuracy between 8 layer QMLP and DeepQMLP models was below 1\%. This empirical study shows that both QMLP and DeepQMLP architectures are trainable and under ideal scenarios (noiseless), the DeepQMLP model performance is at par with the QMLP models. In the following section, we show that the DeepQMLP models may show greater noise resilience during inference due to shallower circuits.

\section{Evaluation} \label{results}

In this section, we evaluate the trainability of the proposed models through empirical studies on various datasets. We train models with a varying number of parametric circuit layers to investigate their impact on performance. Additionally, we compare the performance of QMLP and DeepQMLP during inference under varying degrees of noise and demonstrate that DeepQMLP is more noise resilient.

\begin{figure} 
 \begin{center}
    \includegraphics[width=0.48\textwidth]{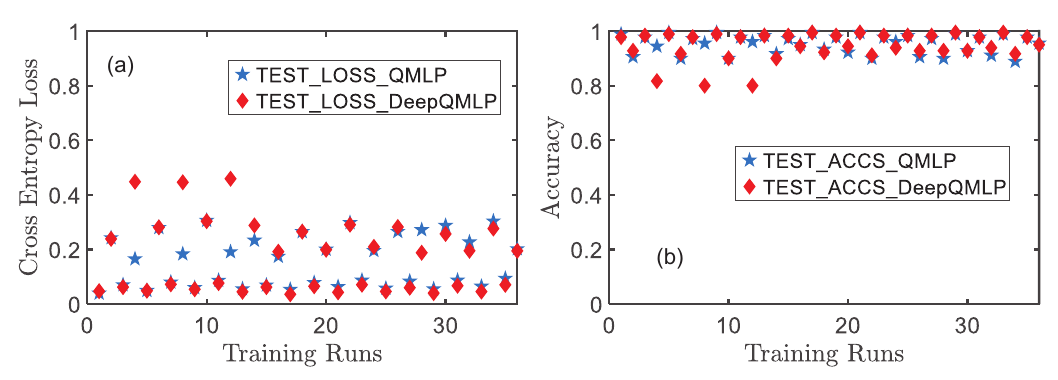}
 \end{center}
 \vspace{-4mm}
\caption{(a) Loss and (b) accuracy of the trained QMLP and DeepQMLP models (4, 6, and 8 parametric layers) on test data (synthetic datasets).}
\label{fig:generalize}
\vspace{-4mm}
\end{figure}

\begin{figure*} 
 \begin{center}
    \includegraphics[width=0.95\textwidth]{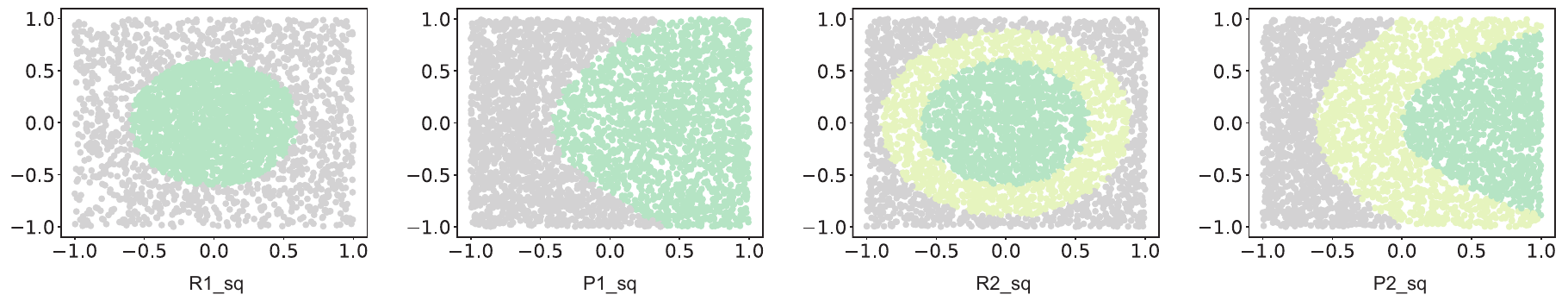}
 \end{center}
\vspace{-4mm}
\caption{Four synthetic datasets used in this work. The colors denote association to different classes. R1\_sq and P1\_sq datasets have 2 classes. R2\_sq and P2\_sq datasets contain 3 classes.} 
\label{fig:synthetic}
\vspace{-4mm}
\end{figure*}

\begin{figure*} 
 \begin{center}
    \includegraphics[width=0.95\textwidth]{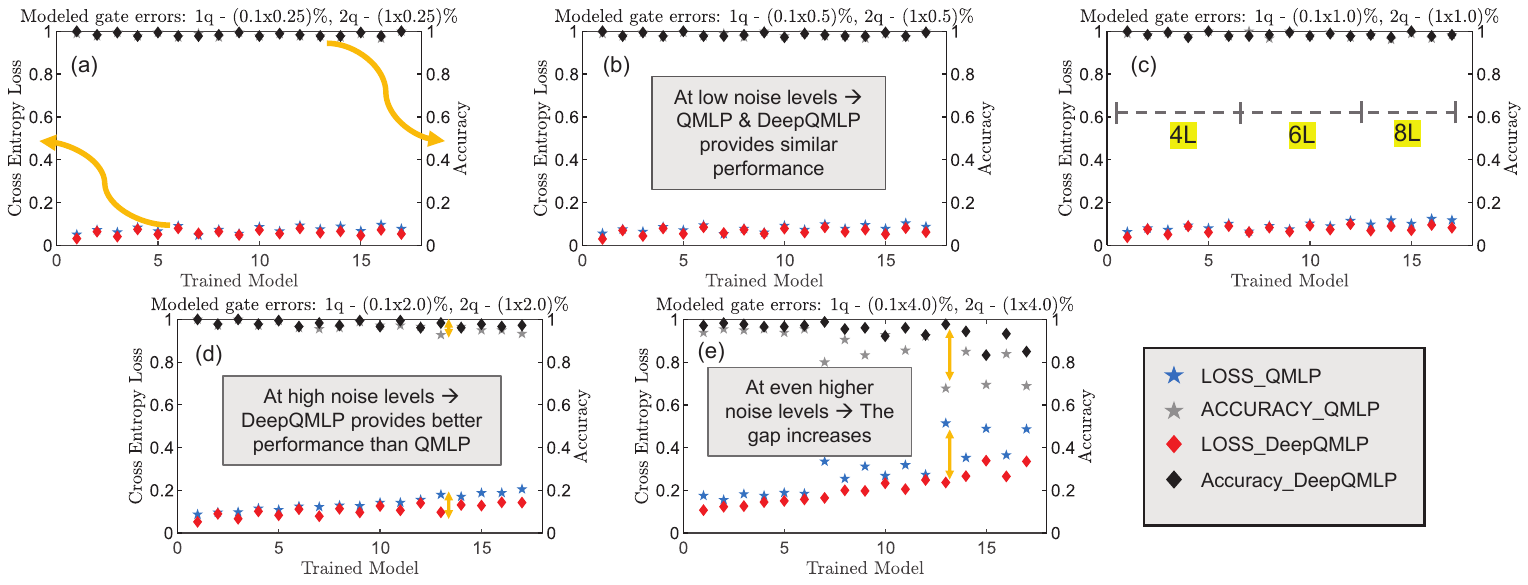}
 \end{center}
 \vspace{-4mm}
\caption{Loss and accuracy of the trained QMLP and DeepQMLP models on the synthetic training datasets under varying levels of gate noises (a)-(e) (1q and 2q denote single-qubit and two-qubit gates, respectively). Under lower noise levels, both QMLP and DeepQMLP provide similar loss and accuracy in inference. At larger noise levels, DeepQMLP models show better loss and accuracy over QMLP models. The gap increases at even higher noise levels which indicate greater robustness of DeepQMLP over QMLP.} 
\label{fig:inference}
\vspace{-4mm}
\end{figure*}

\noindent {\bf{Datasets:}} Apart from the `Iris' dataset, we use four other synthetic datasets - R1\_sq, P1\_sq, R2\_sq, and P2\_sq with non-linear decision boundaries for classification as shown in Fig. \ref{fig:synthetic}. All these synthetic datasets have 2 features (both features are continuous variables varying between -1 and +1). R1\_sq and P1\_sq data samples are divided into two classes while R2\_sq and P2\_sq datasets have 3 classes each (denoted by different colors in Fig. \ref{fig:synthetic}). A total of 180 samples from each dataset are randomly picked for the training purpose. To keep the datasets balanced, we choose 90 samples/class for R1\_sq and P1\_sq and 60 samples/class for R2\_sq and P2\_sq. 

\noindent {\bf{Framework and Setup:}} We develop a Python framework to implement the networks in this study using Pennylane, Pytorch, and Qiskit frameworks \cite{bergholm2018pennylane, paszke2019pytorch, cross2018ibm}. We use the Pennylane framework to model the quantum circuits and Pytorch to model and train the hybrid network. Qiskit is used to perform all the simulations of quantum circuits under noise. In all the training runs, we use an initial learning rate of 0.5 and train the models using mini-batch gradient descent with a batch size of 30. We use the Adagrad optimizer which updates the learning rate for each trainable parameter across multiple epochs based on their update rates in the previous epochs. This enables us to start all the training with a relatively higher learning rate of 0.5. We restrict the training to 50 epochs. 

We use the noisy device emulator in Qiskit (FakeMelbourne) to gauge the impact of various noise sources (gate errors, decoherence, and measurement errors) on the performance of trained QML models. We also use depolarizing noise channels to simulate gate errors in isolation for one of the studies \cite{ash2019qure}. The Pauli-Z expectation values at the output of the PQC are calculated analytically in noise-less simulation. Noisy expectation values are calculated from 10000 samples. Interested readers can look into the noise simulation of the Qiskit framework for further details \cite{cross2018ibm}.


\noindent {\bf{Trainability of QMLP and DeepQMLP:}} In the previous section, we have shown the training cost/accuracy curves of QMLP and DeepQMLP with varying number of layers to classify the `Iris' dataset. Here, we pick the synthetic datasets and train them using QMLP and DeepQMLP architectures with 4, 6, and 8 parametric layers. Note that, a 4 Layer QMLP model has the same number of trainable parameters as a 4 Layer DeepQMLP model (2 parametric layers per hidden quantum layer) in this work. A total of 4x3x2 or 24 models are picked to further study the trainability of QMLP and DeepQMLP. Each model is trained separately from 3 different random initializations of the parameters/weights. A total of 24x3 or 72 training runs are performed on the synthetic datasets. The cross-entropy loss and accuracy over the entire datasets after 50 epochs of training are shown in Fig. \ref{fig:training}(c) and (d). The first 24 (12 for QMLP and 12 for DeepQMLP) data points (from left to right) are for 4 layer models. The next 24 data points are for the 6 layer models, and the last 24 data points are for the 8 layer models. Note that, apart from 3 different runs (DeepQMLP models with 4 layers - 2 in each hidden layers), all the remaining training runs ended with cross-entropy loss less than 0.2 and accuracy close to 100\%. These results indicate that the proposed QMLP and DeepQMLP models are trainable. The 3 outliers in the study may be the result of poor initialization of the parameters.

\noindent {\bf{Generalization of the training:}} To investigate the generalization capability of the trained models, we pick a further 180 samples from the synthetic datasets (none of them were in the training set) and perform inference with all 72 trained models. The results are shown in Fig. \ref{fig:generalize}. Note that, the loss remained smaller than 0.25, and the accuracy remained over 90\% (except three).

\noindent {\bf{QMLP vs. DeepQMLP:}} 
We perform inference of the trained models with the training data under simulated noise. For a fair comparison, we pick 34 models (17 for QMLP and 17 for DeepQMLP) from the 72 trained models discussed above. In these 34 models, the difference in loss and accuracy between the QMLP and the corresponding DeepQMLP model (with the same dataset, the same number of trainable parameters) were on the lower side at the end of the training. 

We model the single-qubit and two-qubit gate errors using depolarizing noise channels \cite{ash2019qure}. Note that, in this simulation, all the single-qubit gates in the circuit have a constant error probability (nominal value assumed to be 0.1\%). Similarly, all the two-qubit gates have a nominal error probability of 1\%. Later we swept these error probability values with the following scaling factors: 0.25, 0.5, 1.0, 2.0, and 4.0. Note that, the nominal 0.1\% and 1\% error probabilities of the single-qubit and two-qubit gate errors are at par with the reported noise levels of the current generation of IBM quantum computers. We avoided other quantum device architectural constraints (e.g., limited connectivity \cite{ash2019qure}) to avoid unnecessary complexity in the comparison. The results are shown in Fig. \ref{fig:inference}(a)-(e).

For lower error values, both QMLP and DeepQMLP models showed similar loss and accuracy over the training data (for scaling factors of 0.25, 0.5, and 1.0 in Fig. \ref{fig:inference}(a)-(c)). This is expected because, at lower noise levels, the output states of the quantum circuits (both in QMLP and DeepQMLP) are not far from the ideal. However, at a higher noise level (scaling factor 2), the DeepQMLP model showed lower loss and higher accuracy over the QMLP models as shown in Fig. \ref{fig:inference}(d) (indicated by the yellow double arrows). At these noise levels, both QMLP and DeepQMLP hidden quantum layers produce erroneous output states that are far from the ideal. However, the shallow-depth DeepQMLP hidden layer outputs are less erroneous compared to the QMLP hidden layers because of a smaller number of gates per hidden layer. Therefore, the overall network performance is less affected by gate noises in DeepQMLP models. On average, DeepQMLP showed 20.86\% lower loss and 1.12\%, higher accuracy over the QMLP models at 2x scaling of the noise. The gap increased further at 4x scaling of the noise as evident in Fig. \ref{fig:inference}(e). On average, DeepQMLP models showed 25.3\% lower loss and 7.93\% higher accuracy over the QMLP models at 4x noise scaling. This study indicates a greater noise resilience of DeepQMLP over QMLP.

\section{Conclusion} \label{conclusion}

We present two new quantum-classical hybrid neural network architectures QMLP and DeepQMLP for classical data classification. We show the trainability of these models through empirical studies on 5 different datasets and 78 training runs. We also show that the trained models generalize well. The DeepQMLP model shows greater noise resilience over the QMLP models (up to 25.3\% lower loss and 7.93\% higher accuracy). These architectures provide new directions to develop large-scale machine learning applications.

\textbf{Acknowledgements:} The work is supported in parts by NSF (CNS-1722557, CNS-2129675, CCF-1718474, OIA-2040667, DGE-1723687, DGE-1821766, and DGE-2113839) and seed grants from Penn State ICDS and Huck Institute of the Life Sciences. 

\bibliographystyle{IEEEtran}
\bibliography{ref}

\end{document}